\documentclass[aps, prl, twocolumn]{revtex4-1}
\usepackage{amsmath}    % need for subequations
\usepackage{graphicx}   % need for figures
\usepackage{verbatim}   % useful for program listings
\usepackage{color}      % use if color is used in text
\usepackage{subfigure}  % use for side-by-side figures
\usepackage{hyperref}   % use for hypertext links, including those to external documents and URLs
\begin{document}

\title{Topological magneto-optics in the non-coplanar antiferromagnet Co$_{1/3}$NbS$_2$: Imaging and writing chiral magnetic domains}

\author{E. Kirstein$^{1}$, H. Park$^{2,3}$, I. Martin$^{2}$, J. F. Mitchell$^{2}$, N. Ghimire$^{4}$, S.~A. Crooker$^{1}$}
\affiliation{$^1$National High Magnetic Field Laboratory, Los Alamos National Lab, Los Alamos, NM 87545, USA}
\affiliation{$^2$Materials Science Division, Argonne National Lab, Lemont, IL 60439, USA}
\affiliation{$^3$Department of Physics, University of Illinois Chicago, Chicago, IL 60607, USA}
\affiliation{$^4$Department of Physics, Notre Dame University, Notre Dame, IN 46556, USA}

%\affiliation{$^*$Authors to whom correspondence should be addressed: crooker@lanl.gov}
%\date{\today}

\begin{abstract}
Despite its tiny net magnetization, the antiferromagnetic (AFM) van der Waals material Co$_{1/3}$NbS$_2$ exhibits a large transverse Hall conductivity $\sigma_{xy}$ even at zero applied magnetic field, which arises, as recently shown, from the topological nature of its non-coplanar ``tetrahedral'' AFM order. This triple-\textbf{q} magnetic order can be regarded as the short-lengthscale limit of a magnetic skyrmion lattice, and has an intrinsic spin chirality.  Here we show, using optical wavelengths spanning the ultraviolet to infrared (400-1000~nm), that magnetic circular dichroism (MCD) provides an incisive optical probe of the topological AFM order in Co$_{1/3}$NbS$_2$.  Measurements as a continuous function of photon energy are directly compared with first-principles calculations, revealing the influence of the underlying quantum geometry on optical conductivity.  Leveraging the power and flexibility of optical methods, we use scanning MCD microscopy to directly image chiral AFM domains, and demonstrate writing of chiral AFM domains.
\end{abstract}

\maketitle
%\section{\label{sec:level1}Introduction\protect\\}
The anomalous Hall effect and related magneto-optical phenomena such as the Kerr and Faraday effects have historically  been associated with materials possessing a net magnetization (e.g., ferromagnets). However, recent discoveries of \textit{anti}ferromagnetic (AFM) materials that also elicit anomalously large Hall \cite{Kubler:2014, Chen:2014, Surgers:2014, Nakatsuji:2015, Nayak:2016} and magneto-optical responses \cite{Feng:2015, Higo:2018, Balk:2019} -- despite their vanishing net magnetization -- have revitalized and focused considerable efforts \cite{Shindou:2001, Martin:2008, Kato:2010, Sivadas:2016, Suzuki:2017, Yanagi:2023, Zhang:2020} to explore noncollinear AFM spin configurations whose lower symmetry permits the appearance of off-diagonal (i.e., Hall) elements of the conductivity tensor, $\sigma_{xy}(\omega)$. Similar to the case for ferromagnets, many of these complex noncollinear AFM orders also require spin-orbit coupling and concomitant spin splitting of the underlying bands, in order to manifest non-zero $\sigma_{xy}$. 

In parallel, a fascinating class of materials -- usually ferromagnetic -- are those exhibiting \textit{non-coplanar} spin textures such as magnetic skyrmions, which can generate $\sigma_{xy}$ solely via  the topological nature of their inherent spin chirality (i.e., without requiring spin-orbit coupling). These ``topological Hall effects'' (THE) \cite{Ye:1999, Ohgushi:2000, Taguchi:2001, Bruno:2004,  Neubauer:2009, Kurumaji:2019, WangTHEreview:2022, Takagi:2023, Park:2023} and related topological magneto-optical (TMO) phenomena \cite{Feng:2020, Schilberth:2022, Kato:2023, Li:2024, Cai:2024} can be regarded as emerging from the geometric (Berry) phase acquired by an electron traversing a chiral spin texture in real space, which acts as a fictitous magnetic field that couples to the electron's orbital degrees of freedom. Disentangling these topological contributions to $\sigma_{xy}(\omega)$ from those arising from background ferromagnetism and spin-orbit coupling, and understanding how spin textures in real space manifest themselves in the Berry curvature of the underlying bands in momentum space, is a topic of active debate and study \cite{Zhang:2020, Verma:2022, Kimbell:2022, Schilberth:2022, Wang:2023, Savary:2025}.  To date, investigations of TMO effects in magnetic materials are at an early stage, with very recent studies reporting signatures of topological Kerr rotation and circular dichroism induced by skyrmion spin textures \cite{Li:2024, Kato:2023, Cai:2024} in ferromagnets such as Gd$_2$PdSi$_3$, CrVI$_6$, and Fe$_3$GeTe$_2$, and suggestions that TMO phenomena may exhibit quantization at low (THz) frequencies \cite{Feng:2020}.

In this context, it would be highly desirable to explore TMO phenomena in fully-compensated  AFM materials that exhibit non-coplanar spin chirality, and are free from background ferromagnetism. The intercalated transition-metal disulfide Co$_{1/3}$NbS$_2$ has recently emerged as an ideal platform for such investigations. It is a van der Waals AFM hosting layers of spins on 2D triangular lattices -- an archetypal template for geometrically frustrated magnetism on which many forms of complex magnetic order can emerge. In 2018, electrical transport studies unexpectedly revealed a very large and hysteretic Hall effect in Co$_{1/3}$NbS$_2$, below its antiferromagnetic ordering temperature, $T_N \approx 28$~K, despite its tiny net magnetization \cite{Ghimire:2018}. This result was particularly surprising because early neutron diffraction studies had reported simple collinear AFM order in Co$_{1/3}$NbS$_2$ \cite{Parkin:1983}, for which non-zero Hall conductivity $\sigma_{xy}$ is nominally forbidden by symmetry. This discovery and its subsequent confirmation \cite{Tenasini:2020, Mangelsen:2021} therefore strongly suggested a far more interesting magnetic structure than originally appreciated. And indeed, very recent neutron diffraction \cite{Takagi:2023} and M\"ossbauer spectroscopy \cite{Dong:2024}, together with first-principles calculations \cite{Park:2022, Heinonen:2022}, indicate that the magnetic ground state of Co$_{1/3}$NbS$_2$ has non-coplanar ``tetrahedral'' AFM order.  Crucially, this triple-\textbf{q} AFM order, depicted in Fig. 1a and discussed in detail below, can be regarded as the short-wavelength limit of a magnetic skyrmion lattice \cite{Shindou:2001, Martin:2008, Kato:2010, Takagi:2023, Park:2023}, wherein Hall conductivity $\sigma_{xy}$ arises from the geometrical phase acquired by electron spins moving within and following the topologically non-trivial chiral spin texture. Because this chiral texture is periodic over just a few unit cells, its impact on the underlying Berry curvature of the reconstructed electronic bands can readily be captured by ab initio calculations.

Importantly, because $\sigma_{xy}(\omega)$ is frequency-dependent, it can manifest not only in electrical transport measurements where $\omega \approx 0$, but also at high (tera- and petahertz) frequencies, thereby providing a route toward all-optical studies of chiral antiferromagnetism via TMO effects. At a given optical frequency, $\sigma_{xy}(\omega)$ is governed by the quantum geometric properties of the electronic bands involved in the interband optical transitions. These bands acquire Berry curvature due to the chiral nature of the AFM spin order, as calculated and discussed below.

Here, using methods for magnetic circular dichroism (MCD), we show that the intrinsic spin chirality of the tetrahedral AFM spin order in Co$_{1/3}$NbS$_2$ generates a large TMO response across the entire visible spectrum.  The data closely follow the topological Hall effects observed in electrical transport studies, thereby validating optical methods as an incisive tool for probing chiral spin order in Co$_{1/3}$NbS$_2$ and related topological AFMs. We investigate the dependence of the MCD on photon energy, and compare the results to first-principles calculations of $\sigma_{xx}(\omega)$ and $\sigma_{xy}(\omega)$. Furthermore, we use scanning MCD microscopy to directly image the micron-scale chiral AFM domain structures that spontaneously form upon cooling below $T_N$.  And finally, we demonstrate writing of chiral AFM domain patterns using light, in analogy with magneto-optical data storage in ferromagnets.  These results lay groundwork for spectrally, spatially, and temporally-resolved studies of topological magneto-optical phenomena in chiral antiferromagnets. 

\begin{figure}[tbp]
\centering
\includegraphics[width=.99\columnwidth]{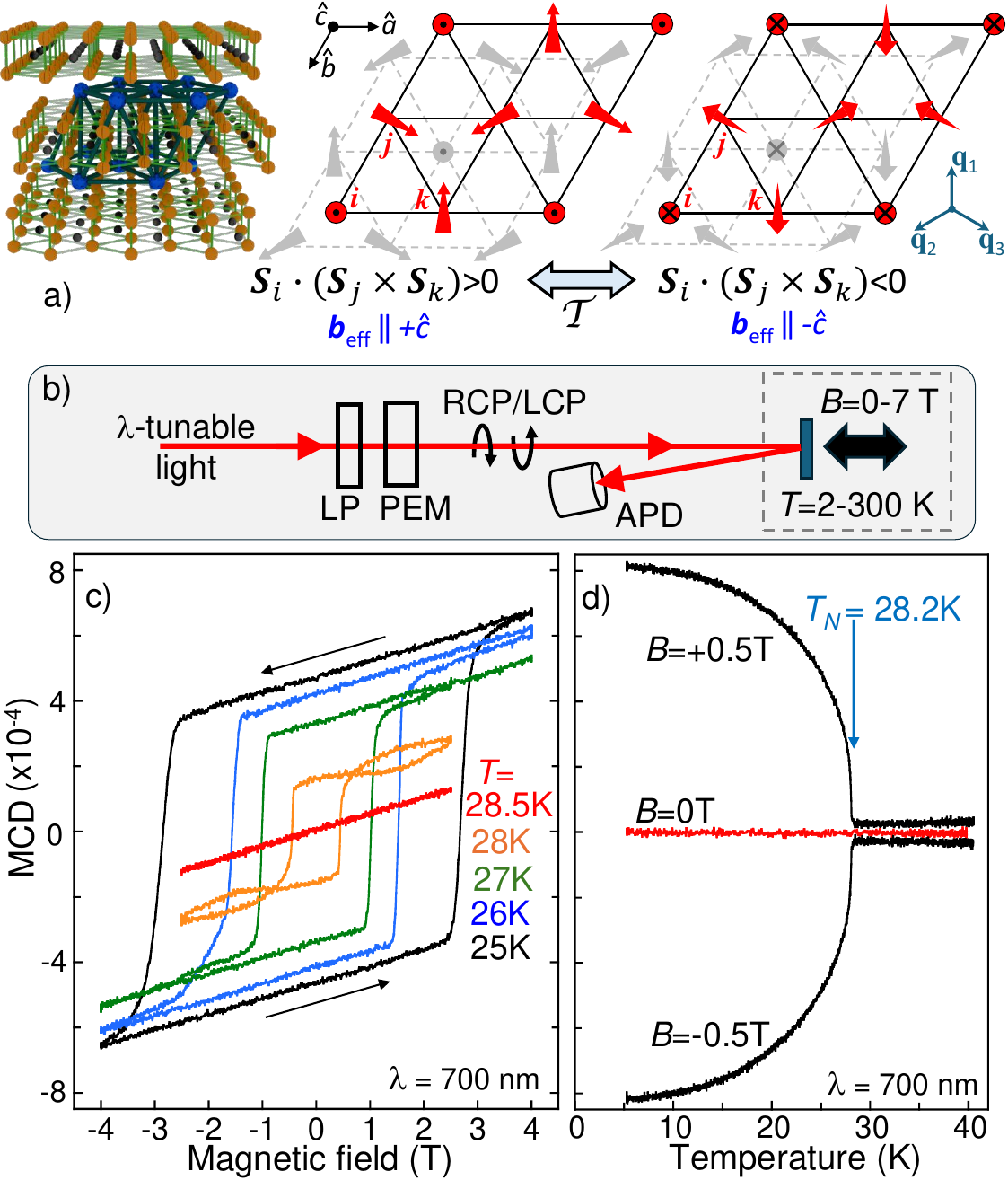}
%includegraphics[totalheight=14cm]{Fig1.pdf}
\caption{(a) Crystal structure and magnetic unit cell of Co$_{1/3}$NbS$_2$. Below its AFM ordering temperature $T_N$, the Co spins exhibit non-coplanar tetrahedral order. Time-reversed ($\mathcal{T}$) spin configurations exhibit opposite scalar spin chirality, $\chi_{ijk} = \mathbf{S}_i \cdot (\mathbf{S}_j \times \mathbf{S}_k)$, from which $\mathbf{b}_{\textrm{eff}}$, $\sigma_{xy}(\omega)$, and TMO effects arise. $\odot$ / $\otimes$ represent spins pointing out of /into the page, and spins with small/large arrowheads are canted into/out of the page. Grey spins depict Co on a neighboring (lower) 2D layer, so that the all-in/all-out tetrahedral spin configuration can be readily recognized. b) The MCD experiment: Wavelength-tunable light is linearly polarized (LP) and modulated between right- and left-circular polarization (RCP/LCP) by a photoelastic modulator (PEM), then reflected at near-normal incidence from the sample and detected by an avalanche photodiode (APD). c) MCD versus $B$ at different temperatures, showing the emergence of large and hysteretic MCD below $T_N \approx 28.2$~K, similar to the large and hysteretic Hall conductivity observed in transport studies (where $\omega \approx 0$) \cite{Ghimire:2018}. d) MCD as Co$_{1/3}$NbS$_2$ is cooled in $B=\pm 0.5$~T and 0~T. The growth of the MCD below $T_N$ tracks the growth of the AFM order parameter $\chi_{ijk}$.}  
\end{figure}

The crystal and magnetic structures of Co$_{1/3}$NbS$_2$ are shown in Fig. 1a. Co$_{1/3}$NbS$_2$ has non-centrosymmetric hexagonal crystal space group $P6_3 22$, wherein layers of Co atoms, intercalated between NbS$_2$ monolayers, form 2D triangular lattices with ABAB stacking. Recent experimental \cite{Takagi:2023, Dong:2024} and theoretical studies \cite{Park:2022, Heinonen:2022},  point to a non-coplanar triple-\textbf{q} AFM ground state wherein the Co spins are oriented ``all-in'' (towards) and ``all-out'' (away from) the centers of the 3D tetrahedral network defined by the Co atoms, as depicted. Tetrahedral spin order $\mathbf{S(r)}$ can arise in triangular lattice AFMs from small effective four-spin interactions between conduction electrons and localized moments \cite{Martin:2008, Kato:2010}, and can be most simply expressed as a superposition of three single-\textbf{q} stripe AFM orders, e.g., 
\begin{equation}
\mathbf{S(r)} = (\Delta_1~\textrm{cos}~\mathbf{q}_1 \cdot \mathbf{r}, \Delta_2~\textrm{cos}~\mathbf{q}_2 \cdot \mathbf{r}, \Delta_3~\textrm{cos}~\mathbf{q}_3 \cdot \mathbf{r}),
\end{equation}
with equal amplitudes $|\Delta_i|$, and in-plane wavevectors $\mathbf{q}_i$ equal to half the reciprocal lattice vectors and rotated $\pm 120^\circ$ with respect to each other. This gives a 4-sublattice magnetic order with spins directed along (1,1,1), (1,-1,-1), (-1,1,-1), and (-1,-1,1). [Note that to correlate Eq. (1) with Fig. 1a, the (1,1,1) direction can be taken as pointing out-of-page].

This tetrahedral spin texture has vanishing net magnetization, but nonetheless breaks all the symmetries (such as the product of time-reversal and translation) that otherwise preclude the appearance of spontaneous Hall conductivity $\sigma_{xy}(\omega)$ in the absence of an applied magnetic field $B$. Importantly, circulating around \textit{any} triangular plaquette (e.g., clockwise from sites $i \rightarrow j \rightarrow k \rightarrow i$; see Fig. 1a), the three non-coplanar spins $\mathbf{S}_{i,j,k}$ subtend a solid angle equal to one-quarter of a sphere and have a non-zero scalar spin chirality $ \mathbf{S}_i \cdot (\mathbf{S}_j \times \mathbf{S}_k)$ \cite{Martin:2008, Kato:2010, Zhang:2020, Takagi:2023, Park:2023}. Thus, the eight plaquettes within each 2D magnetic unit cell can be considered to contain two skyrmions. Moreover, every 2D plane has the same sign and magnitude of chirality.  Electrons moving in this chiral spin background acquire a geometrical Berry phase that acts similarly to the Aharonov-Bohm phase induced by a real magnetic field $B$, leading to an out-of-plane effective field $\mathbf{b}_{\textrm{eff}}$, a non-zero $\sigma_{xy}(\omega)$, and the emergence of TMO phenomena. More formally, the periodic chiral spin texture generates Berry curvature of the reconstructed electronic bands in momentum space, even without spin-orbit coupling, so that $\sigma_{xy}(\omega) \neq 0$.  Applied out-of-plane $B$ can switch between time-reversed spin configurations (see Fig. 1a), giving opposite spin chirality, $\mathbf{b}_{\textrm{eff}}$, and $\sigma_{xy}(\omega)$. 

Figure 1b depicts the MCD experiment. Bulk single crystals of Co$_{1/3}$NbS$_2$ were grown by chemical vapor transport and characterized as described in Refs. \cite{Ghimire:2018, Tenasini:2020}. Their high quality is evinced by their large $T_N \approx$28\,K and sharp magnetic switching (shown below). They were freshly exfoliated and mounted within the variable-temperature insert of a 7~T superconducting magnet. Wavelength-tunable probe light between 400-1000 nm, derived from a xenon lamp filtered through a spectrometer, was modulated between right- and left-circular polarization at 50~kHz. The probe light was weakly focused on the sample surface ($\sim$1~mm spot size), and the reflected light was detected by a photodiode. MCD measures the normalized difference between the reflected intensities, $(I_R - I_L)/(I_R + I_L)$, and is closely related to the Kerr ellipticity (i.e., the imaginary part of the complex Kerr angle) \cite{Cai:2024, Li:2024}.  For MCD imaging studies requiring high spatial resolution, the samples were instead mounted in a small helium optical cryostat, and 650~nm light from a superluminescent diode was focused by a microscope objective to a small ($\sim 1~\mu$m) spot that was raster-scanned across the sample.  

Figure 1c shows MCD measurements of Co$_{1/3}$NbS$_2$ vs. $B$, at different temperatures $T$. At 28.5~K (just above $T_N$) and at all higher temperatures, the MCD signal vanishes at $B$=0, and exhibits only a featureless linear background.  However at 28.0~K (just below $T_N$), a hysteresis loop emerges and the MCD signal becomes non-zero, even at $B$=0, after ramping down from large $|B|$.  Upon further cooling by just a few degrees, the hysteretic MCD grows in amplitude and the coercive field $B_c$ that is required to switch between time-reversed chiral spin configurations (\textit{cf.} Fig. 1a) rapidly increases. Below 24~K, $B_c$ exceeds our maximum 7~T field. Figure 1d shows the onset and emergence of the large positive (negative) MCD signal as Co$_{1/3}$NbS$_2$ is cooled in a small $B$=+0.5~T (-0.5~T), which effectively poles the sample to a specific AFM chirality. Note, however, that the MCD signal remains tiny when cooled in $B$=0. As shown later, the sample does spontaneously form micron-sized AFM domains of positive and negative spin chirality below $T_N$, but the large (1~mm) size of the probe beam in these studies averages over many domains. 

The MCD signals shown in Figs. 1c and 1d, measured using red light at 700~nm, closely follow the $B$- and $T$-dependent Hall effects reported in electrical transport studies of Co$_{1/3}$NbS$_2$ \cite{Ghimire:2018}. These data therefore validate TMO methods as an incisive probe of chiral AFM order parameter in Co$_{1/3}$NbS$_2$, and suggest that the the full power and flexibility of optical techniques can be brought to bear on the study of topological antiferromagnets, including spectrally-, spatially-, and temporally-resolved studies.  

\begin{figure}[tbp]
\centering
\includegraphics[width=0.90\columnwidth]{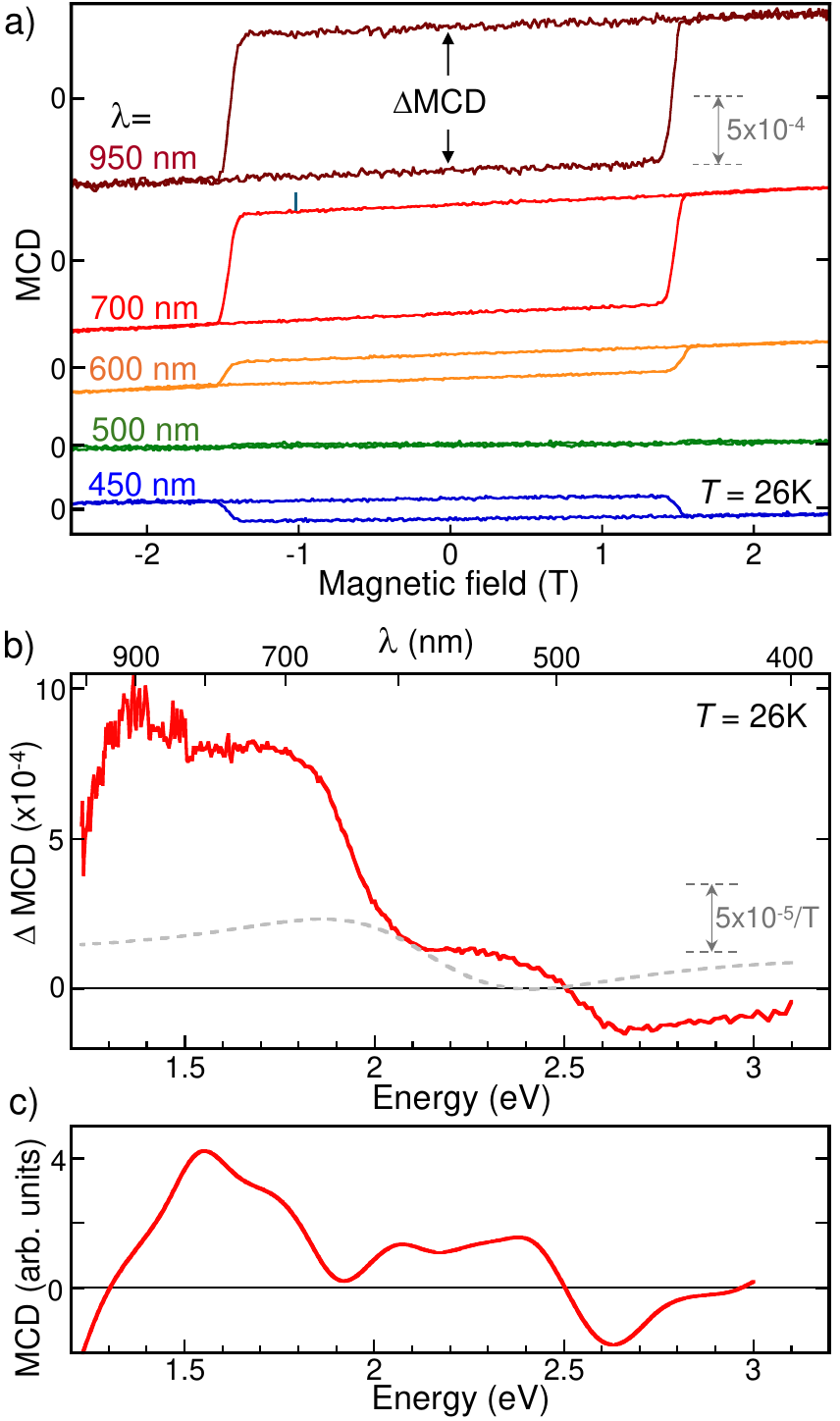}
\caption{a) MCD vs. $B$ at $T$=26~K, using different wavelengths (photon energies) of probe light. All curves are plotted on the same vertical scale (shown), but offset for clarity. The amplitude and sign of the hysteresis loops vary with wavelength, as does the slope of the linear background. b) Dependence of the amplitude of the hysteresis loop, $\Delta$MCD, on photon energy (i.e., the TMO response elicited solely by chiral AFM order). Dashed grey line shows the slope of the linear background (in units of MCD/T; smoothed). c) First-principles calculation of the energy-dependent MCD arising from non-coplanar tetrahedral AFM order in Co$_{1/3}$NbS$_2$.}
\end{figure}

As a first step, we measure the dependence of the MCD response on optical wavelengths spanning the near-infrared (1000~nm) to the ultraviolet (400~nm). These results provide detailed information about the frequency dependence of $\sigma_{xy}(\omega)$ -- a quantity often calculated from first principles using density-functional theory (DFT) \cite{Feng:2015, Schilberth:2022, Feng:2020, Sangalli:2012, Yao:2004}.  Figure 2a shows MCD($B$) using different probe wavelengths. Notably, both the amplitude \textit{and sign} of the hysteresis loop -- that is, the TMO response elicited by the chiral AFM order -- varies with wavelength, inverting sign at $\approx$500~nm and becoming quite large in the near-infrared. In addition, the background linear slope also varies. These measured spectral dependencies are explicitly shown as a continuous function of photon energy in Fig. 2b. 

This spectral dependence is inextricably linked to the underlying magnetic order and encodes information about the interband optical transitions that contribute to the anomalous and topological Hall effects at higher energies, providing a spectral TMO ``fingerprint'' for different magnetic materials.  For direct comparison with first-principles theory, we calculated the optical conductivities $\sigma_{xx,xy}(\omega)$ and estimated the MCD ($\sim$ Im[$\sigma_{xy}(\omega)$]/Re[$\sigma_{xx}(\omega)$]) arising from tetrahedral spin order in Co$_{1/3}$NbS$_2$ and its influence on the Berry curvature of the bands in momentum space; see Supplementary Figs. S1 and S2. Spin-orbit coupling was \textit{not} included -- the calculated response is due to the chirality of the noncoplanar spin texture alone. The overall agreement with experiment is quite satisfactory; namely, both data and theory show a steep rise at the lowest energies ($\sim$1.3~eV) leading to a broad maximum near 1.5~eV, which then falls to a local minimum near 2~eV followed by a broad plateau before inverting sign at $\sim$2.5~eV.  Moreover, we note that the measured data trend towards the additional sign inversions predicted at the short- and long-wavelength limits of our detection range. (Note that at any given probe energy, TMO effects are generally non-zero but are not quantized, being dominated by the quantum geometry of bands only in the relevant energy window.) As a control study, calculations of collinear AFM order was confirmed to give no MCD response. Calculation details, momentum-space Berry curvatures, and conductivities are given in the Supplemental Material. 

\begin{figure}[tbp]
\centering
\includegraphics[width=.95\columnwidth]{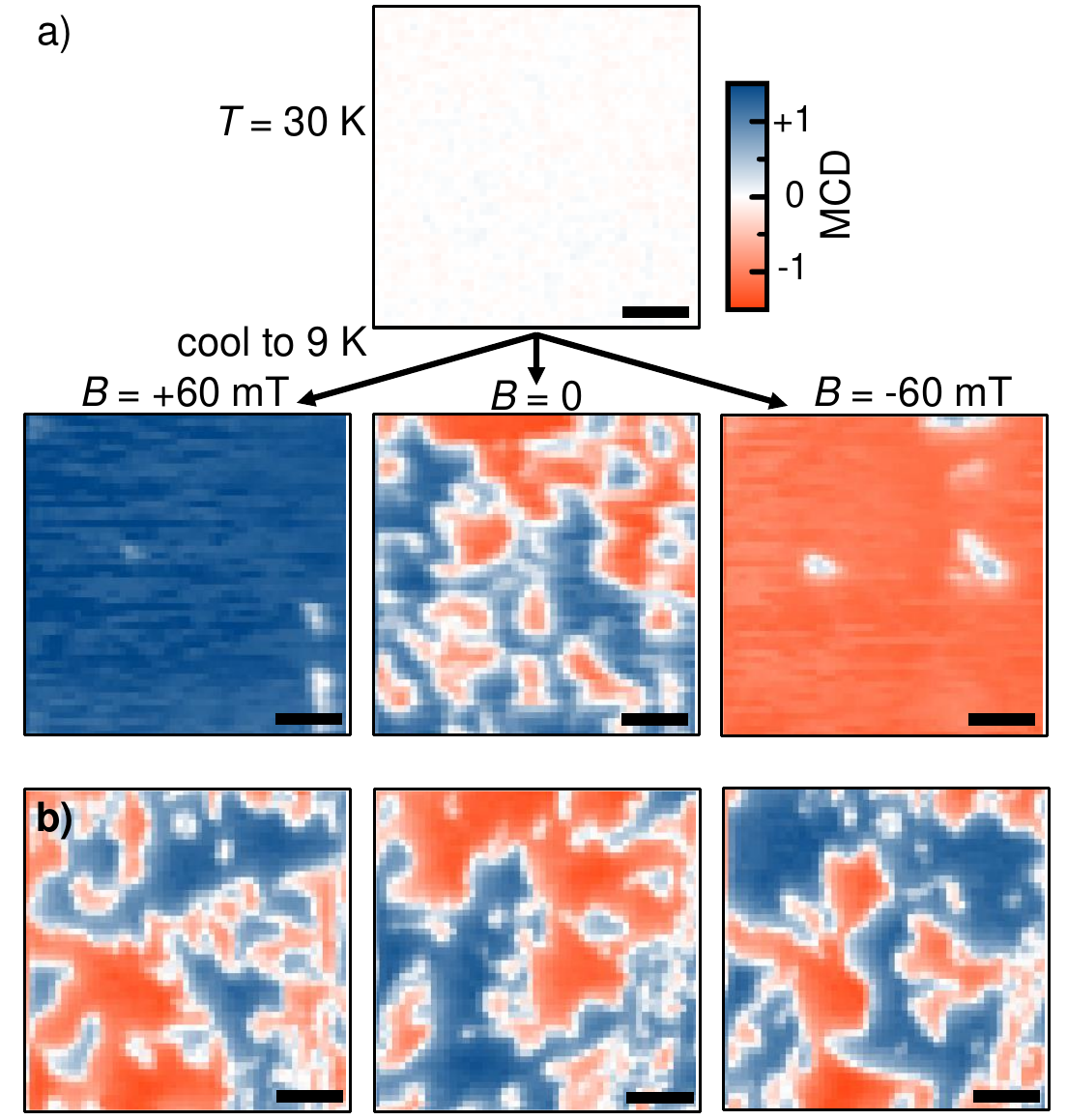}
\caption{50$\times$50 $\mu$m MCD images of Co$_{1/3}$NbS$_2$, at a) $T$=30K ($>T_N$), and after cooling to 9~K ($<T_N$) in $B$=+60~mT, 0~mT, and -60~mT respectively. Chiral AFM domains form spontaneously when cooled in $B$=0. Scale bars are 10 $\mu$m. b) Images of chiral AFM domains at $T$=7.5~K, in the same area, following three thermal cycles to 30~K at $B$=0.} 
\end{figure}

The temperature-dependent MCD shown in Fig. 1d demonstrates that Co$_{1/3}$NbS$_2$ can be readily poled to a particular handedness of the chiral AFM state by cooling in small $\pm B$.  However, cooling in $B=0$ showed no net signal. To investigate whether small chiral AFM domains are forming spontaneously at zero field, we constructed a scanning MCD microscope for imaging studies (1 $\mu$m spatial resolution). The images in Fig. 3 show that when Co$_{1/3}$NbS$_2$ is cooled below $T_N$ even in very weak $B$=$\pm 60$~mT, the sample becomes almost uniformly polarized with positive or negative chirality. However, cooling in $B$=0 results in spontaneous symmetry breaking and formation of micron-scale AFM domains with opposite chirality (\textit{cf} Fig. 1a). The colorscale shows that within each domain, the MCD signal is as large as that from a uniformly-poled sample. The characteristic domain size is $\approx$5 $\mu$m, similar to a related report \cite{Gu:2023}, and the domain pattern varies from cooldown to cooldown, indicating an absence of pinning centers that favor a particular AFM chirality or localize a domain wall (moreover, note that the inherent \textit{crystalline} chirality of Co$_{1/3}$NbS$_2$ does not appear to play a role in determining the AFM domain structure). Looking forward, it will be interesting to study how these spontaneous chiral domain patterns evolve as Co$_{1/3}$NbS$_2$ is exfoliated down to the atomically-thin limit, to study the role of dimensionality and inter-layer spin interactions. 

\begin{figure}[tbp]
\centering
\includegraphics[width=0.95\columnwidth]{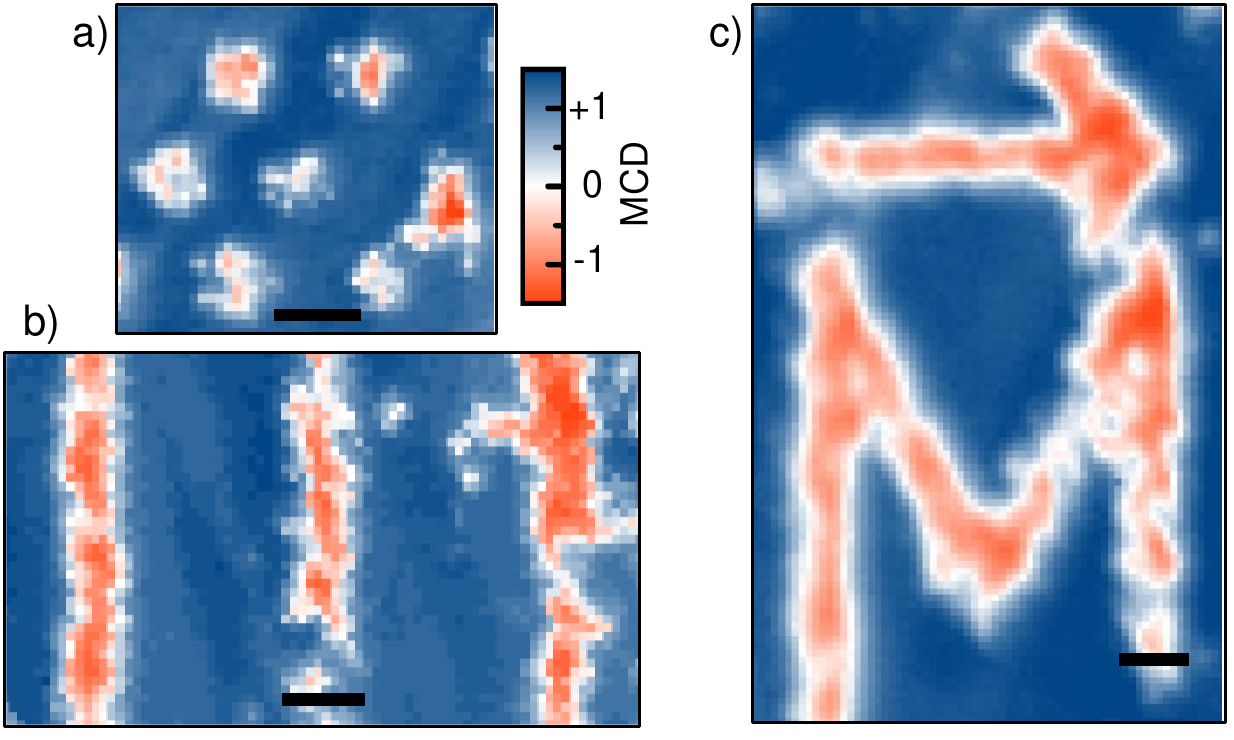}
\caption{Writing AFM domains with opposite spin chirality in Co$_{1/3}$NbS$_2$ using light (see main text).  The MCD images show examples of chiral AFM domain patterns comprising a) a triangular arrangement of local domains, b) vertical stripes, and c) the National High Magnetic Field Lab's logo, ``$\vec{\textrm{M}}$''. Scale bars are 25~$\mu$m.}
\end{figure}

In addition to passive imaging studies, the high spatial resolution of optical microscopy also provides the means to use focused light to deliberately engineer material properties on small length scales. The capability to not only read but also to write magnetic domains is essential for applications in magnetic information storage and spintronics. In this context, antiferromagnets are widely recognized as promising platforms \cite{Jungwirth:2016, Baltz:2018, Smejkal:2018, DalDin:2024, Han:2023, Cheong:2023}, owing to their fast switching speeds and relative immunity to stray magnetic fields. To this end, in Fig. 4 we demonstrate the ability to locally write chiral AFM domains in Co$_{1/3}$NbS$_2$, using a thermally-assisted approach analogous to that used in commercial magneto-optical storage devices based on ferromagnets \cite{Immink:1984}.  Here, a Co$_{1/3}$NbS$_2$ sample is cooled below $T_N$ in a small $B$ ($+60$~mT) to pole it into a positive chiral AFM state. Then $B$ is reversed (-60~mT), which causes no change because $B_c$ is much larger (\textit{cf.} Fig 1c).  A weak continuous-wave laser (1~mW, 633~nm) is then used to briefly and locally heat a small region of the sample above $T_N$, which, upon re-cooling, is poled by $B$ to the negative chiral AFM state. The resulting patterns are then imaged.  Figure 4 shows patterns comprising a triangular arrangement of time-reversed chiral AFM domains, stripes, and more complex patterns. The size of the written domains ($\approx$ 5-8 $\mu$m) is larger than the excitation spot ($\approx$ 1 $\mu$m), and is likely determined by the thermal diffusion length in this bulk sample. Warming above $T_N$ erases all patterns. In these studies, we did not find that the optical polarization of the weak writing beam played a role.  Smaller domain features can likely be written with thin-film samples or nm-thick exfoliated layers, or possibly by using ultrafast optical pulses from pulsed lasers \cite{Kimel:2004}. In the future, we anticipate that electrical transport studies through engineered patterns and arrays of topological AFM domains may be of particular interest. %We note that a related thermo-magnetic methods was used recently to demonstrate optical writing of domains in the noncollinear AFM Mn$_3$Sn \cite{Reichlova:2019}. 

In summary, these studies demonstrate the power and utility of optical methods to investigate topologically non-trivial antiferromagnets, via TMO phenomena. Broadband optical spectroscopy based on circular dichroism and/or Kerr effects can provide a spectral TMO ``fingerprint'' of different AFM materials and their underlying magnetic orders, against which first-principles calculations can be compared and benchmarked. The ability to image and also write chiral AFM domains using light opens up a rich playground for detailed studies of AFM domain formation and dynamics, analogous to related advances in ferromagnets over the past several decades.

We thank P. Park, C. Batista, F. Ronning, and A. Balk for helpful discussions. E.K. was supported by the Los Alamos LDRD program, and S.C. was supported by the U.S. Department of Energy (US DOE) Quantum Science Center. H.P. and I.M. (theory and simulations), and N.G. and J.F.M. (crystal growth and characterization) were supported by the US DOE Office of Science, Basic Energy Sciences, Materials Sciences and Engineering Division. The National High Magnetic Field Lab is supported by National Science Foundation (NSF) DMR-1644779, the State of Florida, and the US DOE.

\renewcommand{\thefigure}{S\arabic{figure}}
\setcounter{figure}{0}

\newpage

\begin{figure*}
\center
\section{Supplemental Material}

\includegraphics[width=1.8\columnwidth]{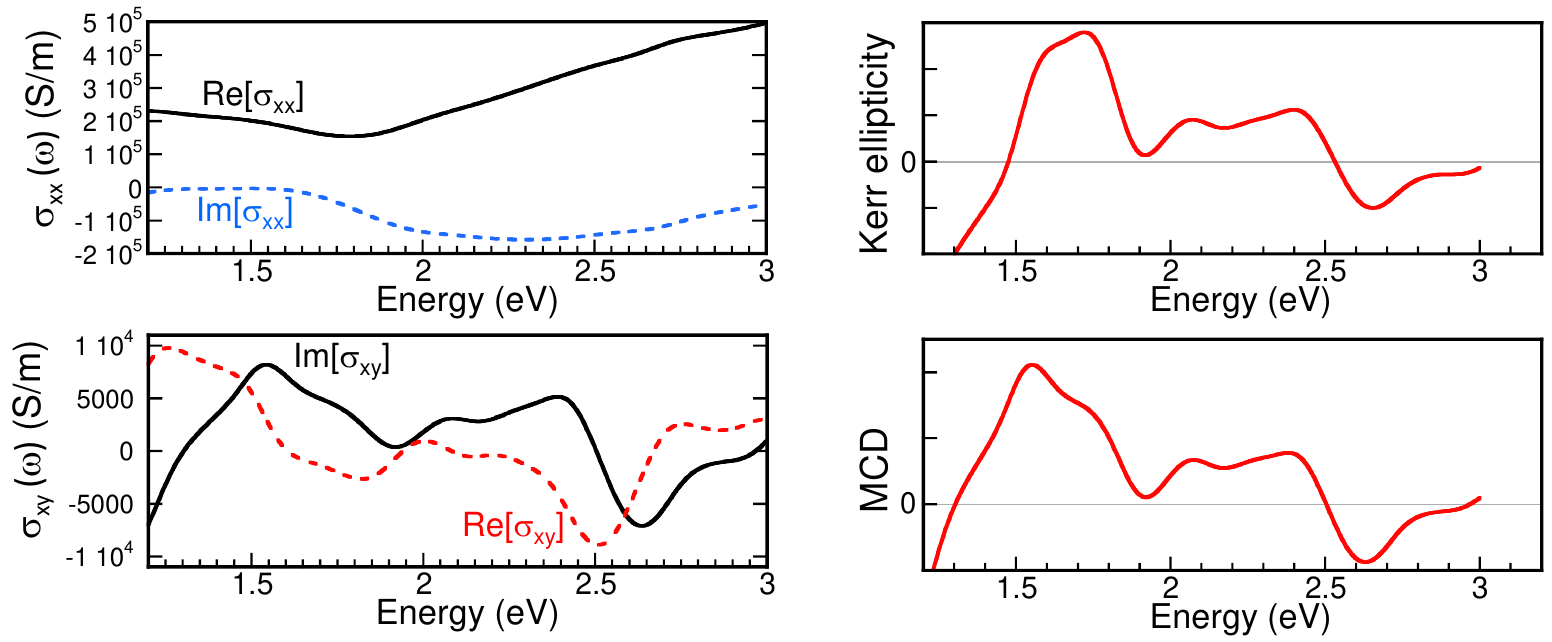}
\caption{\textbf{First-principles calculations:} To calculate the energy-dependent optical conductivities and topological magneto-optical (TMO) response from the non-coplanar tetrahedral antiferromagnetic (AFM) spin order in Co$_{1/3}$NbS$_2$, we adopt the Vienna Ab-initio Simulation Package (VASP)~\cite{Kresse:1996, Kresse:1999} code for density functional theory (DFT) calculations using the experimental structure of Co$_{1/3}$NbS$_2$ measured from X-ray diffraction~\cite{Anzenhofer:1970}. The Perdew-Burke-Ernzerhof (PBE)~\cite{Burke:1997} functional is used for the exchange-correlation functional with the plane-wave energy cut-off for 400~eV and a 14x14x4 k-point grid. For the magnetic band structure and the dynamical Hall conductivity calculations, we first construct the tight-binding Hamiltonian based on the non-magnetic primitive cell using both Co \textit{d} and Nb \textit{d} Wannier orbitals by adopting the Wannier90 code~\cite{Marzari:2012}. The magnetic unit cell is constructed as the 2x2x1 supercell from the primitive cell. Then, the spin-exchange interaction is incorporated into the Hamiltonian and treated using the Hartree-Fock approximation following previous approaches~\cite{Park:2022}. \textbf{Left:} The dynamical Hall conductivities $\sigma_{xx}(\omega)$ and $\sigma_{xy}(\omega)$ are computed based on the magnetic band structure by adopting the Wannier-Berri package~\cite{Tsirkin:2021}. For convergence, we used a 10x10x8 k-mesh with 10 recursive refinement iterations and the smoothing of the Fermi functions at temperature $T$=10~K. \textbf{Right:} The calculated Kerr ellipticity $\epsilon_K(\omega)$ (i.e., the imaginary part of the complex Kerr angle $\theta_K + i\epsilon_K = -\sigma_{xy}/[\sigma_{xx} \sqrt{1 + i(4\pi/\omega)\sigma_{xx}}~]$) \cite{Feng:2015, Schilberth:2022}.  Shown below for comparison is the calculated MCD from the main text (estimated as $\sim$ Im[$\sigma_{xy}$]/Re[$\sigma_{xx}$]; the trends are similar except that the zero-crossing at low energy occurs at $\sim$1.5~eV instead of $\sim$1.3~eV. }
\label{S1}
\end{figure*}

\newpage

\begin{figure*}
\center
\includegraphics[width=1.9\columnwidth]{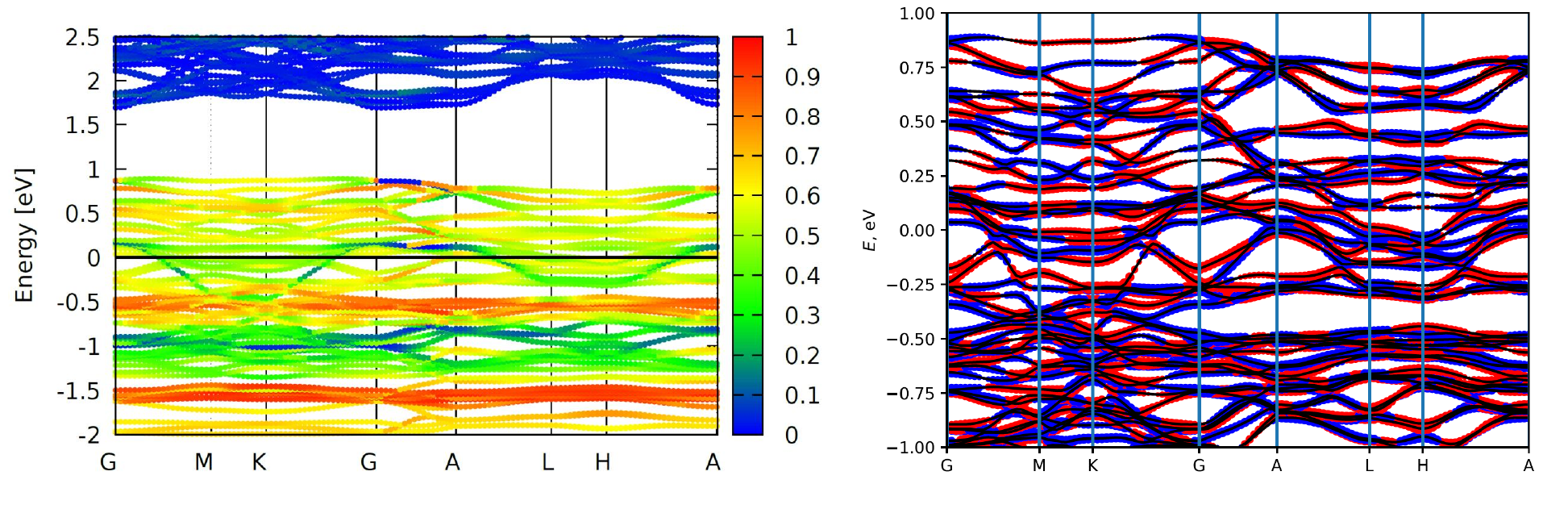}
\caption{(a) The orbital-projected spin-polarized band structure in Co$_{1/3}$NbS$_2$. The red (blue) color indicates the projection to the Co \textit{d} (Nb \textit{d}) characters. (b) The Berry-curvature projected band structure in Co$_{1/3}$NbS$_2$. The red (blue) color means the + (-) sign of the Berry curvature and the dot size indicates the magnitude of the Berry curvature. The Berry curvature of the reconstructed bands in momentum space is generated by the chiral tetrahedral AFM spin texture that forms the magnetic unit cell shown in Fig. 1a of the main text.}
\label{S2}
\end{figure*}

\end{document}